\documentclass[twocolumn,showpacs,preprintnumbers,amsmath,amssymb]{revtex4}

\usepackage{graphicx}
\usepackage{dcolumn}
\usepackage{bm}

\begin{document}
\title{ Hall plateaus at magic angles in bismuth beyond the quantum limit}
\author{Beno\^{\i}t Fauqu\'e$^{1}$, Huan Yang$^{1}$, Ilya Sheikin$^{2}$, Luis Balicas$^{3}$,
Jean-Paul Issi$^{4}$ and Kamran Behnia$^{1}$} \affiliation{(1)LPEM
(UPMC-CNRS), ESPCI, Paris, France\\ (2) Grenoble High Magnetic Field
Laboratory (CNRS), Grenoble , France\\ (3) National High Magnetic
Field Laboratory, Tallahassee, FSU, Florida, USA\\ (4)CERMIN,
Universit\'e Catholique de Louvain, Louvain-la-Neuve, Belgium}

\date {Feb. 9, 2009}

\begin{abstract}

We present a study of the angular dependence of the resistivity
tensor up to 35 T in elemental bismuth complemented by torque
magnetometry measurements in a similar configuration.  For at least
two particular field orientations a few degrees off the trigonal
axis, the Hall resistivity was found to become field-independent
within experimental resolution in a finite field window
corresponding to a field which is roughly three times the frequency
of quantum oscillations. The Hall plateaus rapidly vanish as the
field is tilted off theses magic angles. We identify two distinct
particularities of these specific orientations, which may play a
role in the emergence of the Hall plateaus.
\end{abstract}


\maketitle

The quantum limit is attained when the magnetic field is strong
enough to confine electrons to their lowest Landau level. Beyond
this limit, an interacting two-dimensional electron gas can display
the Fractional Quantum Hall Effect (FQHE)\cite{chakraborty}. In
three dimensions on the other hand\cite{halperin}, the fate of the
electron gas pushed to this ultraquantum regime is barely explored.
Because of its low carrier concentration, elemental
bismuth\cite{edelman} provides a unique opportunity to attain the
extreme quantum limit in a bulk metal with laboratory magnetic
fields. Recent studies on bismuth has uncovered a rich but poorly
understood physics beyond the quantum limit\cite{behnia1,li}. One
central question is to determine if the band picture, which treats
electrons as non-interacting entities, remains valid in such an
extreme limit, where the interactions and their associated
instabilities are enhanced\cite{halperin} and the dimensionality is
reduced\cite{yakovenko,biagini}.

A first study\cite{behnia1} of high-field Nernst and Hall
coefficients in bismuth resolved unexpected anomalies at fields
exceeding 9 T for a field roughly oriented along the trigonal axis.
In this configuration, transport properties and their quantum
oscillations are dominated by the hole-like ellipsoid of the Fermi
surface\cite{bompadre,behnia2}. Since the quantum limit of these
carriers occur at 9 T, the detected anomalies were attributed to
interacting hole-like quasi-particles at fractional filling
factors\cite{behnia1}. Following this observation, a  study of
torque magnetometry\cite{li} detected the quantum oscillations of
the three electron pockets and their angular variation. In addition
to the anomalies caused by the passage of successive Landau levels,
this study resolved a field scale with a sharp angular variation and
identified it as a phase transition of the quasi-particles of the
electron pocket, which, in contrast to holes, present a Dirac
spectrum\cite{wolff}. The link between these two sets of observation
remained unclear. These experimental results initiated new
theoretical investigations regarding the possible occurrence of FQHE
in a bulk system\cite{burnell} as well as the high-field electronic
spectrum of bismuth\cite{alicea,sharlai}.

\begin{figure}
  \resizebox{!}{0.5\textwidth}{\includegraphics{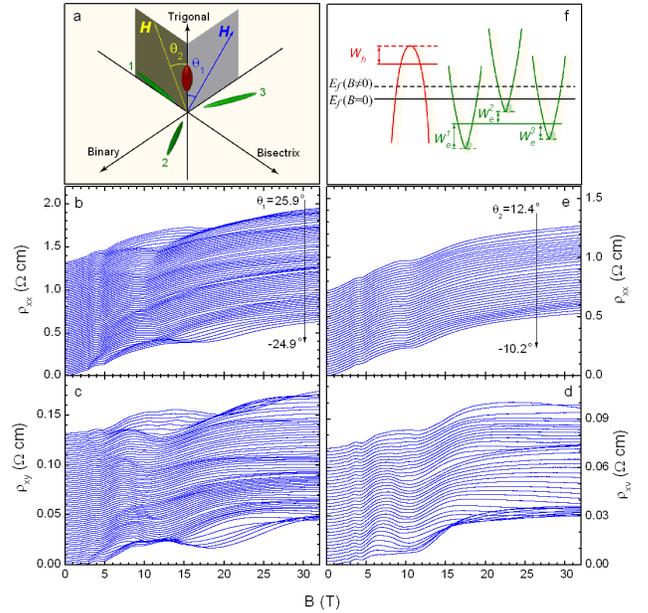}}
\caption{\label{Fig1} a) The Fermi surface of bismuth. The magnetic
field was applied along an orientation tilted off the trigonal axis
by sweeping either $\theta_{1}$ or $\theta_{2}$. Panels b and c (d
and e) present the Hall and resistivity data obtained in the first
(second) case at T= 1 K. Curves are shifted for clarity.  f)
Schematic view of the band structure. In presence of magnetic field
along an arbitrary direction, the Fermi level, the top of the hole
band and the bottom of the electron bands shift to new positions
(dotted lines) in order to maintain charge neutrality. }
\end{figure}

Here we present a study of Hall and longitudinal resistivities in
presence of a strong rotatable magnetic field.  These transport
measurements were preceded by a study of torque magnetometry in the
same configuration, which confirms the observations reported by
Li\emph{ et al.}\cite{li} and provide supplementary insight to the
transport data. The results allow us to conclude that: i) The Hall
response is dominated by the carriers of the hole pocket of the bulk
Fermi surface; ii) The contribution of the quasi-particles of the
electron pocket are visible as a perturbation to the overall
conductivity (both longitudinal and transverse) ; iii) There are
particular orientations of magnetic field (dubbed ``magic angles'')
for which the Hall resistivity becomes field-independent in the
vicinity of 20 T. Such a Hall plateau has not been previously
observed in any bulk quasi-isotropic material and its explanation is
a challenge for the one-particle picture. Two distinct features of
these orientations, which may be relevant to the emergence of the
plateaus can be readily identified in our data.

The samples were all cut from a large single crystal of bismuth
several cm long\cite{boxus}. While the Residual Resisitivity Ratio
(RRR= $\rho$(300K)/ $\rho$(4.2K)) of the mother crystal was 300, the
RRR of the tailored samples with a typical thickness of 0.8 mm was
found to be much lower ($\sim$ 100) pointing to a mean-free-path
long enough to be affected by sample dimensions\cite{garcia}.
Resistivity and Hall effect were measured with a standard 6-contact
set-up. Torque magnetometry was measured using a cantilever and a
high-resolution capacitance bridge. The current was applied along
the bisectrix and the magnetic field was tilted off the trigonal
axis either in the (trigonal, bisectrix) or the (trigonal, binary)
plane of the crystal. Two miniature Hall probes were used to
determine the orientation with a relative resolution lower than 0.1
degree, but with an absolute uncertainty of about 1 degree.

Fig. 1 presents the field dependence of longitudinal,  $\rho_{xx}$,
and transversal, $\rho_{xy}$, resistivity. High-field features,
occurring at fields exceeding the quantum limit are particularly
visible in $\rho_{xy}$. They rapidly evolve as the magnetic field is
tilted a few degrees off the trigonal axis. This rapid angular
evolution indicates that the ultraquantum transport anomalies which
were reported in both bismuth\cite{behnia1} and in
Bi$_{0.96}$Sb$_{0.04}$\cite{banerjee} for a magnetic field roughly
along the trigonal axis, are extremely sensitive to the orientation
of the magnetic field.

\begin{figure}
   \resizebox{!}{0.5\textwidth}{\includegraphics{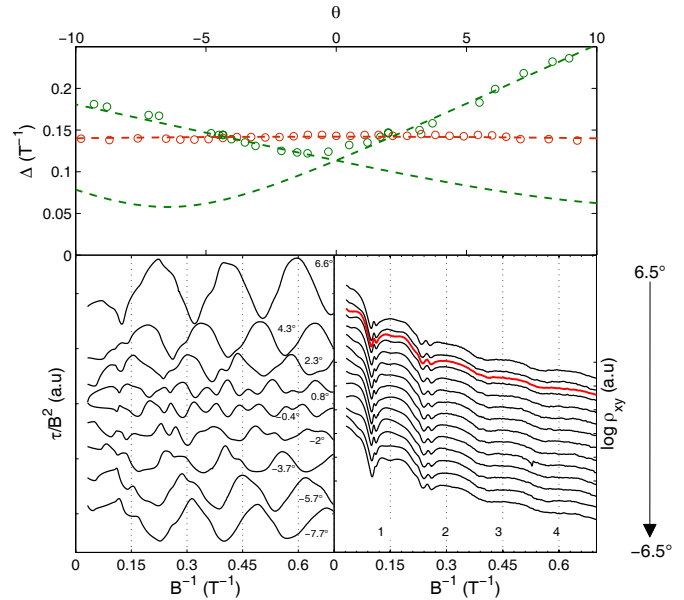}}
\caption{\label{Fig2}  Transverse susceptibility (lower left) and
Hall resistivity (lower right) as a function of the inverse of the
magnetic field for different [$\theta_{2}$] tilt angles slightly off
the trigonal axis. Curves are shifted for clarity. The red Hall
curve corresponds to a magic angle. Upper panel shows the angular
dependence of the period of quantum oscillations seen in the Hall
(red symbols) and torque (green symbols) data. Green (red) dotted
lines correspond to the expected angular dependence of periods for
electron (hole) ellipsoids. }
\end{figure}

The Fermi surface of bismuth consists of a hole ellipsoid and three
electron pockets. In a first approximation, the Hall response of
such a compensated metal, with strictly equal concentration of
electrons and holes, should be zero. A finite signal is expected
when the mobility of one type of carriers exceeds the mobility of
the other. In the case of bismuth, due to a large interband
contribution, the Hall response of the electron pockets is expected
to be non-trivial even in the weak-field limit\cite{fuseya}. In our
experimental configuration (field along trigonal and current along
bisectrix axes), the Hall response of all samples studied was found
to be vanishingly small or slightly negative in the weak-field
limit(below 0.1 T) in agreement with previous reports for this
configuration\cite{mase,friedman,hartman}. Moreover, in all of them
a positive Hall signal emerged when the field exceeded 0.2 T. The
magnitude of this large-field Hall coefficient, $R_{H}$, was found
to be sample dependent but always smaller or of the order of
$\frac{1}{n_{h}e}$ ($n_h=2.7 \times 10^{-17} cm^{-3}$ is the density
of hole-like carriers). The positive sign of the Hall response
indicates that holes dominate the Hall response as a result of
higher hole mobility in this configuration. This conclusion is
supported by the fact that the period of quantum oscillations in all
samples (0.15T$^{-1}$) corresponds to what is reported for the hole
pocket of the \emph{bulk} Fermi surface by the de Haas-van
Alphen\cite{bhargava} and the Shubnikov-de Haas\cite{bompadre}
studies as well as the quantum oscillations of the Nernst
coefficient\cite{behnia2}.

The moderate anisotropy of the hole-like ellipsoid implies that
tilting the magnetic field a few degrees off the trigonal axis does
not significantly modify the period of oscillations. This is indeed
the case as seen in Fig. 2, which compares this feature with the
sharp angular variation of the quantum oscillations of the
transverse magnetic susceptibility ( $ \chi_{\perp}= \tau/B^{2}$,
where  $\tau$ is the magnetic torque). The torque response is
dominated by the more anisotropic and three-fold degenerate electron
pockets whose diamagnetic response is accentuated by their Dirac
dispersion. As the field is tilted, the quantum oscillations of the
torque response rapidly vary as expected for the electron pockets.
As seen in the upper panel, the period of quantum oscillations of
Hall and torque data is in rather good agreement with the expected
periods for electrons and holes.

\begin{figure}
 {\includegraphics{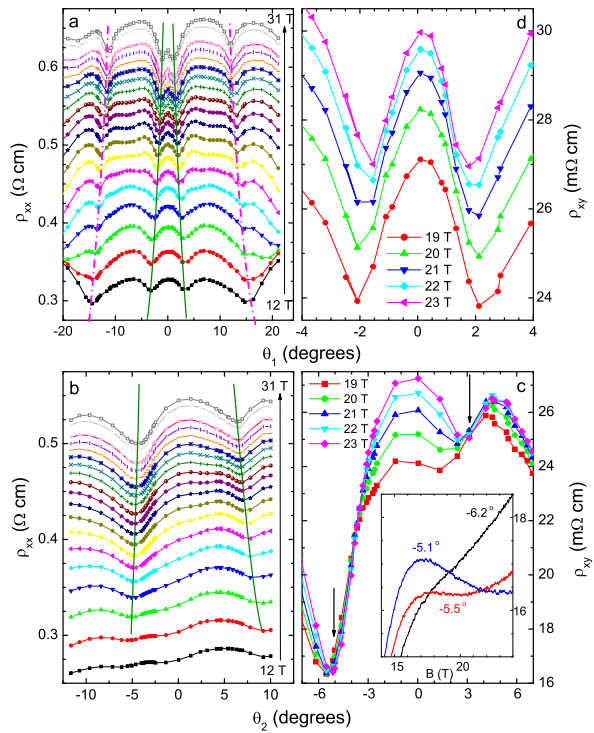}}
\caption{\label{Fig3} a) and b)The angular dependence of the
longitudinal resistivity a function of $\theta_{1}$ and $\theta_{2}$
for different magnetic fields. Lines are guide to lines to follow
the field-dependence of the minima. c) and d) Same for the Hall
resistivity in a restricted field window around 20 T . In panel (c)
arrows indicate magic angles and the inset shows a Hall plateau and
its angular fragility. }
\end{figure}

The contribution of the electron-like carriers to hole-dominated
charge transport can be resolved by putting under scrutiny the
angular dependence of the magnetoresistance, $\rho_{xx}$ as seen in
Fig. 3. The rotating magnetic field generates sharp minima in the
angular dependence of $\rho_{xx}$. When the field was rotated in the
(trigonal,bisectrix) plane, the field dependence of these anomalies
define quasi-vertical field scales in the (B, $\theta_{1}$) plane,
which are symmetrical with respect to the $\theta_{1}=0$ line. The
two central lines lie very close to the field scale reported by Li
and co-workers\cite{li} and identified as a phase transition
involving the electron pockets. Note that this field scale tracks
the $0^{+}_{e}$ Landau levels of two electron pockets according to
calculations \cite{alicea,sharlai}. In this configuration, the
minima in $\rho_{xx}(\theta_{1})$  and in $\rho_{xy}(\theta_{2})$
are concomitant, the Hall response does not present any additional
structure and  does not become field-independent in any finite field
window, at least up to 28 T.

The lower panels of the same figure present the data obtained for
the same crystal with the same contacts for a field rotating in the
(trigonal,bisectrix) plane. Here also minima in
$\rho_{xx}(\theta_{2})$ trace quasi-vertical lines in the (B,
$\theta_{2}$) plane. However, the angular separation between these
lines exceeds what is expected according to the calculated Landau
levels of electrons\cite{sharlai}. More strikingly, the angular
dependence of the Hall response presents an additional structure.
There are two narrow angular windows in which,
$\rho_{xy}(\theta_{2})$ becomes field-independent in the vicinity of
20 T. In other words, there are two orientations for which
$\rho_{xy}$ does not vary with magnetic field in a finite field
window. As seen in the inset, the Hall plateau rapidly vanishes as
the field is tilted a fraction of degree away from these ``magic
angles''.
\begin{figure}
 {\includegraphics{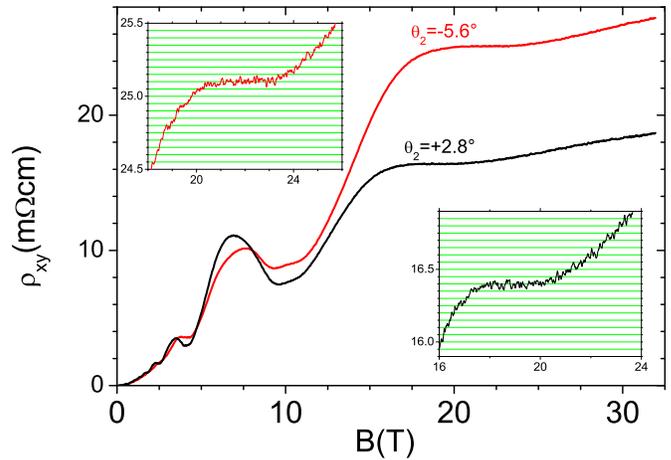}}
\caption{\label{Fig4} The field dependence of $\rho_{xy}$ at magic
angles.  The Insets are zooms on restricted windows with horizontal
lines separated by 0.05m$\Omega$ cm indicating a flatness of 5
10$^{-3}$.}
\end{figure}

Fig.4 presents the field-dependence of $\rho_{xy}$ at two magic
angles. As seen in the inset, that the flatness of the Hall
resistivity is comparable with the experimental noise
($3\times10^{-3}$).  The Hall plateaus centered around 19 T and 21
T, roughly three times the main frequency of the quantum
oscillations, B$_{0}=0.15^{-1}T$. Naively, this corresponds to a
filling factor of 1/3 for holes. However, both the carrier density
and the effective filling factor at high fields and arbitrary angles
could be significantly different from the one estimated from the
low-field spectrum.

In absence of magnetic field, the anisotropy of charge conductivity
in elemental bismuth is less than two. This quasi-isotropy
distinguishes the context of our observation from all cases of Hall
plateaus including the Integer Quantum Hall Effect seen in bulk
layered systems such as the Bechgaard salts\cite{cooper}. Moreover,
in contrast with the non-dissipative behavior expected for an
incompressible quantum Hall fluid, longitudinal resistivity in our
samples remains always finite.
\begin{figure}
 \resizebox{!}{0.58\textwidth} {\includegraphics{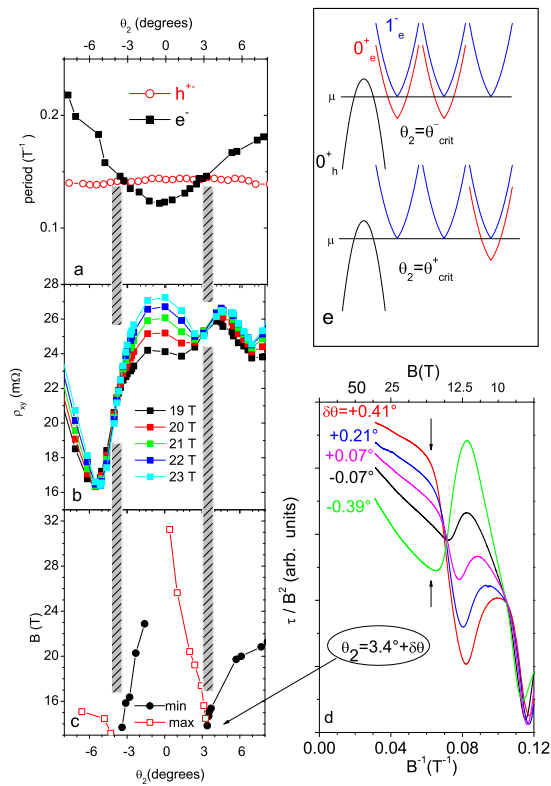}}
\caption{\label{Fig5} : Left panel : Comparison of the angular
dependence of the a) period quantum oscillations, b) $\rho_{xy}$ and
c) high-field torque anomalies. Gray vertical bars mark magic
angles. d) Drastic change of transverse susceptibility near a magic
angle. Arrows indicate anomalies which are tracked in panel c . e)
Schematic representation of Landau levels at two critical angles
possibly corresponding to the experimentally observed magic angles.}
\end{figure}

Finding a credible scenario for a Hall plateau in a bulk system in
presence of the z-axis degeneracy remains a challenge. The presence
of several subsystems adds twists to the problem. As the magnetic
field is swept or rotated, both electrons and holes modify their
zero-field band parameters in order to insure charge
neutrality\cite{smith,vecchi}. The steady and large high-field
magnetostriction\cite{michenaud1} suggests a sizeable field-induced
correction of the carrier density(Fig. 1f), which could
significantly affect the transport properties\cite{michenaud2}. The
restriction of the Hall plateaus to a finite angular and field
window points to a subtle balance of parameters. According to a
recent theoretical work, FQHE can occur in a bulk quasi-isotropic
system only if the electrons re-organize themselves in layers
perpendicular to the magnetic field.  It is tempting to speculate
that the ``magic angles'' correspond to a specific reorganization of
charge distribution fulfilling the required conditions. Transport
measurements along $z$-axis would be helpful for checking this
hypothesis.

Experimentally, the angular separation between the two magic angles
is 8.4 degrees. Two distinct and possibly relevant features of these
two field orientations can be readily identified. The first concerns
both holes and electrons. As seen in 5a, close to the two magic
angles, the periods of quantum oscillations for holes and electrons
become equal. This may be an accident. On the other hand, the
possible commensurability of hole and electron wave-vectors along
the magnetic field for these particular orientations may lead to an
instability paving the way to the emergence of the Hall plateau. The
other specificity of a magic angle concerns solely the electron
pockets. As seen in Fig. 5 (panels c and d), a drastic change in
torque response occurs when the field orientation crosses a magic
angle. According to the theoretical phase diagram\cite{sharlai},
this point corresponds to a simultaneous Landau level crossing of
distinct electron pockets. As detailed above, the Hall plateaus
emerge only when the field rotates in the (trigonal, bisectrix)
plane. Interestingly  only in this configuration there are two
specific field orientations for which the Landau levels of \emph{all
three} electron pockets cross the chemical potential at the same
field (Fig. 5e). This is because when $\theta_{2}$ is swept, two
electron pockets (no 2 and 3 of Fig. 1a) remain degenerate.
According to theoretical calculations, the angular distance between
these two points in the (B, $\theta_{2}$) plane is 8 degrees
\cite{sharlai}. It has been argued that Coulomb interactions are
significantly enhanced whenever a low Landau level crosses the Fermi
level\cite{alicea}. These two particularities of the magic angles
appear as clues to plausible scenarios for a field-independent Hall
response.

We thank J. Alicea, L. Balents, Y. Kopelevich, D. Maslov, G.
Mikitik,  A. J. Millis and Y. Sharlai and V. Yakovenko for useful
discussions. This work is supported by the Agence Nationale de
Recherche as a DELICE project(ANR-08-BLAN-0121-02). Research at the
NHMFL is supported by the NSF, by the State of Florida and by the
DOE.

\end{document}